\documentclass{IET-Conf-Paper}

\begin{document}

\title{UNSUPERVISED DISAGGREGATION OF WATER HEATER LOAD FROM SMART METER DATA PROCESSING}

\author{Thierry Zufferey\ad{1}\corr, Gustavo Valverde\ad{2}, Gabriela Hug\ad{1}}

\address{\add{1}{Power Systems Laboratory, ETH Zurich, Switzerland}
\add{2}{School of Electrical Engineering, UCR, San José, Costa Rica}
\email{thierryz@ethz.ch}}


\begin{abstract}
In the residential sector, electric water heaters are appliances with a relatively high power consumption and a significant thermal inertia, which is particularly suitable for Demand Response (DR) schemes. The success of efficient DR schemes via the control of water heaters presupposes an accurate estimate of their power demand at each instant. Although the load of water heaters is rarely directly measured, a large penetration of Smart Meters (SMs) in distribution grids enables to indirectly infer this information on a large scale via load disaggregation. For that purpose, a considerable number of Non-Intrusive Load Monitoring (NILM) approaches are suggested in the literature. However, they require data streams at a time resolution in the range of one second or higher, which is not realistic for standard SMs. Hence, this paper proposes an unsupervised approach to detect and disaggregate the load profile of water heaters from standard SM data with a time resolution in the minute range. Evaluated on multiple real loads with sub-metering, the proposed approach achieves a Normalized Mean Absolute Error (NMAE) lower than $2\%$ and a precision generally higher than $92\%$ with time resolutions between 5 and 15 minutes.
\end{abstract}

\maketitle

\section{Introduction}

In future distribution grids, Demand Response (DR) is seen as an efficient way to integrate an increasing share of Renewable Energy Sources (RES) as the low controllability of RES can be compensated by the flexibility potential on the demand side. This potential can be exploited via price signals which discourage consumption at high loading conditions and encourages consumption in periods with great energy production. The industrial sector already profits from such time-varying electricity prices. Alternatively, some electric devices could potentially be directly controlled by Distribution System Operators (DSOs) without impacting the comfort of the user, especially in the tertiary and residential sector. The thermal inertia of Thermostatically Controlled Load (TCLs) such as air conditioners, space heaters and electric Water Heaters (WHs) offers great flexibility, which is particularly interesting for DR applications~\cite{Gils,Soder}. In addition, the direct control of TCLs does not require any intervention by the user, in contrast to domestic appliances like washing machines and dishwashers, or even electric vehicles. In this paper, we focus on electric storage WHs which are among the appliances with the greatest flexibility potential and thermal storage~\cite{Pipattanasomporn,Dhulst}. They are usually the largest consumers in a building in terms of rated power, the ramp rate from zero to full rated power is almost instantaneous, and their energy storage potential is available over several hours and can be exploited at any time, as their usage is only marginally impacted by seasonal variations~\cite{Clarke}.

In order to properly design DR schemes, the consumption pattern and the availability of the individual devices under consideration must be estimated. Multiple approaches have been proposed in literature for that purpose. Some studies rely on typical load profiles or on the physical modeling of the electrical behavior of flexible devices, associated with statistical surveys about their consumption share in a certain population~\cite{Ericson, Sajjad}. Nevertheless, these approaches only give a rough and general estimate of the availability at an aggregate level and cannot correctly account for the high diversity and volatility of electrical devices. Recently enabled by the wide-scale roll-out of advanced metering devices, and especially SMs, data-based approaches provide the means for an accurate quantification of the flexibility potential of the customers in a specific distribution system. Sub-metering of the devices of interest gives perfect knowledge of their consumption pattern and of their flexibility potential, as shown by the Belgian LINEAR pilot project~\cite{Dhulst}. Nowadays, smart metering at the building level becomes the norm, but measurements of individual appliances are only feasible in small-scale experiments as their high installation cost inhibits a generalization of sub-metering in distribution grids. Alternatively, load profiles of single appliances can be extracted from power measurements at the building level. Non-Intrusive Load Monitoring (NILM) approaches can detect the so-called signature of a specific device based on its unique steady- or transient-state characteristics~\cite{Mocanu,Kong}. Traditionally based on hidden Markov models or neural networks, they can be supervised (i.e., they require a share of sub-metering data as training data) or unsupervised. However, as far as the authors are concerned, all NILM approaches proposed in literature for WH load disaggregation presuppose a time resolution of at least one second (or much higher for the detection of transient-state characteristics). This is far from the standards of SMs that gather measurements at most with a 1-minute granularity.

Therefore, this paper presents a detection and disaggregation approach of WH load profiles easily applicable in smart distribution grids. The approach is uniquely based on power measurements at the building level and requires a time resolution between 1 and 15 minutes, typical for SMs. Although the small steady-state power variations that build the load signature are smoothed out and not detectable at this time resolution, WHs still distinguish themselves by their high active power consumption and their absence of reactive power consumption. The efficiency of the proposed approach is demonstrated on actual SM measurements at varying time resolutions.

The remainder of this paper is structured as follows. Section~\ref{sec:approach} details the unsupervised detection and disaggregation process of WH loads. In Sec~\ref{sec:case_study}, the proposed approach is tested and evaluated on real measurements. Finally, Sec~\ref{sec:conclusion} concludes the paper and discusses future work.

\section{Load Detection and Disaggregation}\label{sec:approach}

Most TCLs are influenced by the outside temperature, which is for example the case for heat pumps and space heaters that are highly active in winter and barely used in summer. This characteristic can be exploited by disaggregation approaches to compare long periods with high and low TCL activity~\cite{Kouzelis, Kipping}. However, this is not applicable to electric WHs whose energy consumption is relatively constant throughout the year, with a difference in hot water demand of only $20\%$ between the coldest and warmest days of the year in Europe~\cite{Gils}. In this paper, we take advantage of a combination of features which are specific to WHs. Notably, WHs are turned on and off almost instantaneously, consume a relatively high and constant active power during ON periods, but do not consume or produce any reactive power. The proposed disaggregation method processes active power measurements in a first step and reactive power measurements in a second step. Note that both the detection of a WH load and its disaggregation are unsupervised, i.e. only the active and, if possible, the reactive SM power measurements at a building level are required.

\subsection{Detection of Water Heater Load}\label{sec:detection}

\begin{figure}[!t]
\includegraphics[width=1\columnwidth]{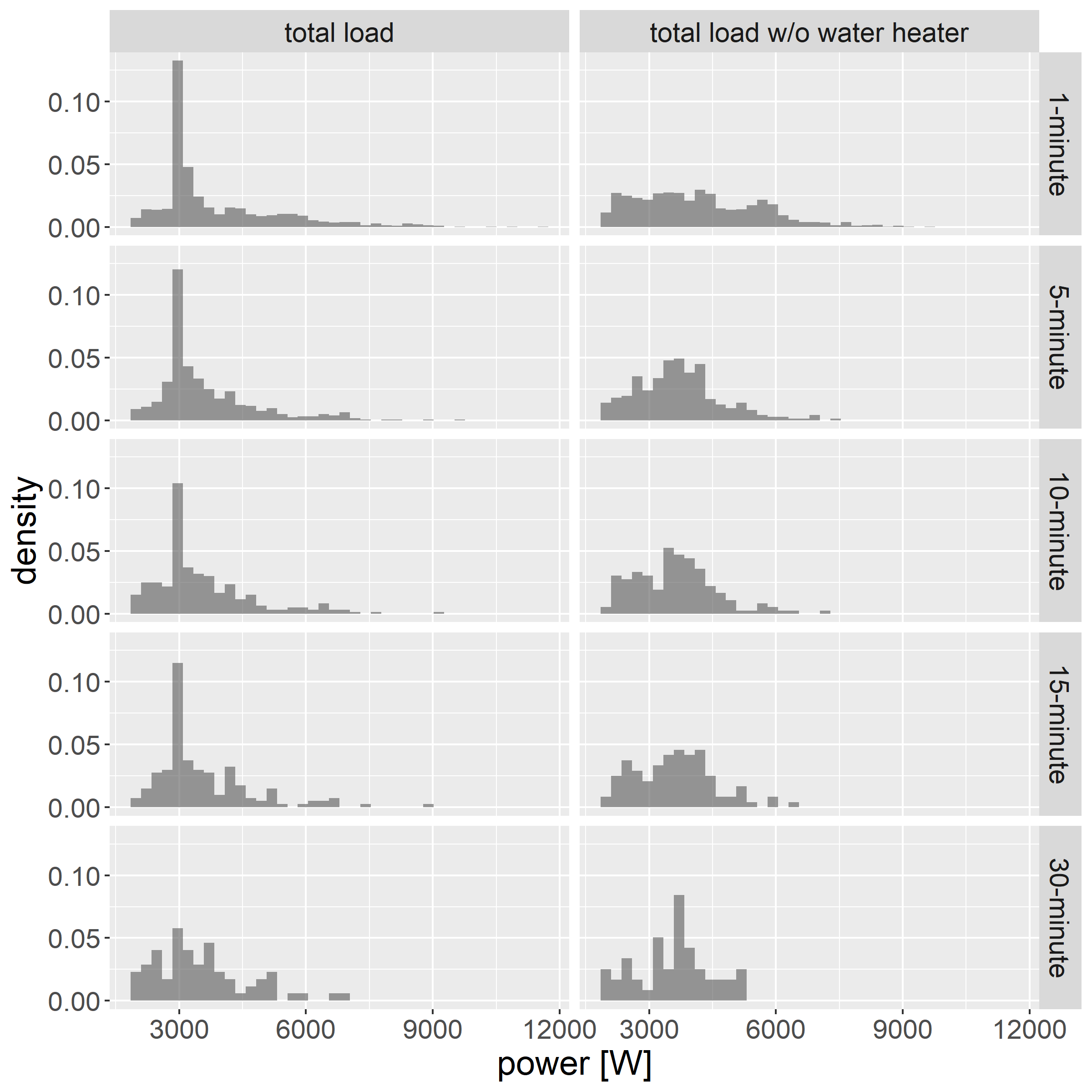}
\caption{Histogram of the total active power consumption at various time resolutions in a representative household with and without a water heater.}
\label{fig:histogram}
\end{figure}

The active power consumption profile of WHs is characterized by large spikes of the same amplitude, corresponding to their rated power. Depending on the hot water demand and the thermal losses, these consumption spikes generally occur several times a day to keep the water temperature in the tank within a certain range. Hence, the presence of electric WHs is visible in the upper part of the active power histogram of a building load, as illustrated by Fig.~\ref{fig:histogram} for a sample household (i.e.,~house 3 in Table~\ref{table:stats}) at various typical SM time resolutions. Indeed, one of the bins exhibits a substantially higher density, which disappears when the WH load is subtracted from the total household load. Note also that the histogram spike induced by the WH load totally disappears when considering 30-minute resolution data, which prevents the detection of the WH. This is due to the smoothing effect of lower time resolutions on SM data, as explained in more detail in Sec.~\ref{sec:evaluation}. In order to filter out the activity of other cyclic ON/OFF devices with a lower rated power (e.g., air conditioners and refrigerators), only power values higher than 2 kW are considered in the histogram~\cite{Pipattanasomporn}. In this paper, the width of the histogram bins is fixed to 200 W. A WH is thus said to be detected if the power histogram contains at least one outlier bin, i.e. the density of at least one bin exceeds the commonly used threshold for outliers~\cite{Tukey}:
\begin{equation}
    d^{\textrm{bin}}_i >= d^{\textrm{bin}}_{Q3} + 1.5\cdot (d^{\textrm{bin}}_{Q3}-d^{\textrm{bin}}_{Q1}),
\end{equation}
where $d^{\textrm{bin}}_i$ is the density of bin $i$ in the total load histogram above 2 kW, and $d^{\textrm{bin}}_{Q1}$ and $d^{\textrm{bin}}_{Q3}$ are the first and third quartiles (i.e., $25^{th}$ and $75^{th}$ percentiles) among the bin density values, respectively.

The rated active power of the detected WHs is estimated as the difference between the power of the highest outlier bin above 2 kW and the power of the first spike in the histogram below 2 kW, corresponding to the base load (e.g., sleep mode of electronic devices). For the representative household of Fig.~\ref{fig:histogram}, a rated power of 3 kW can be correctly estimated.

\subsection{Water Heater Disaggregation Process}\label{sec:disaggreggation}

\begin{figure*}[!t]
\includegraphics[width=1\linewidth]{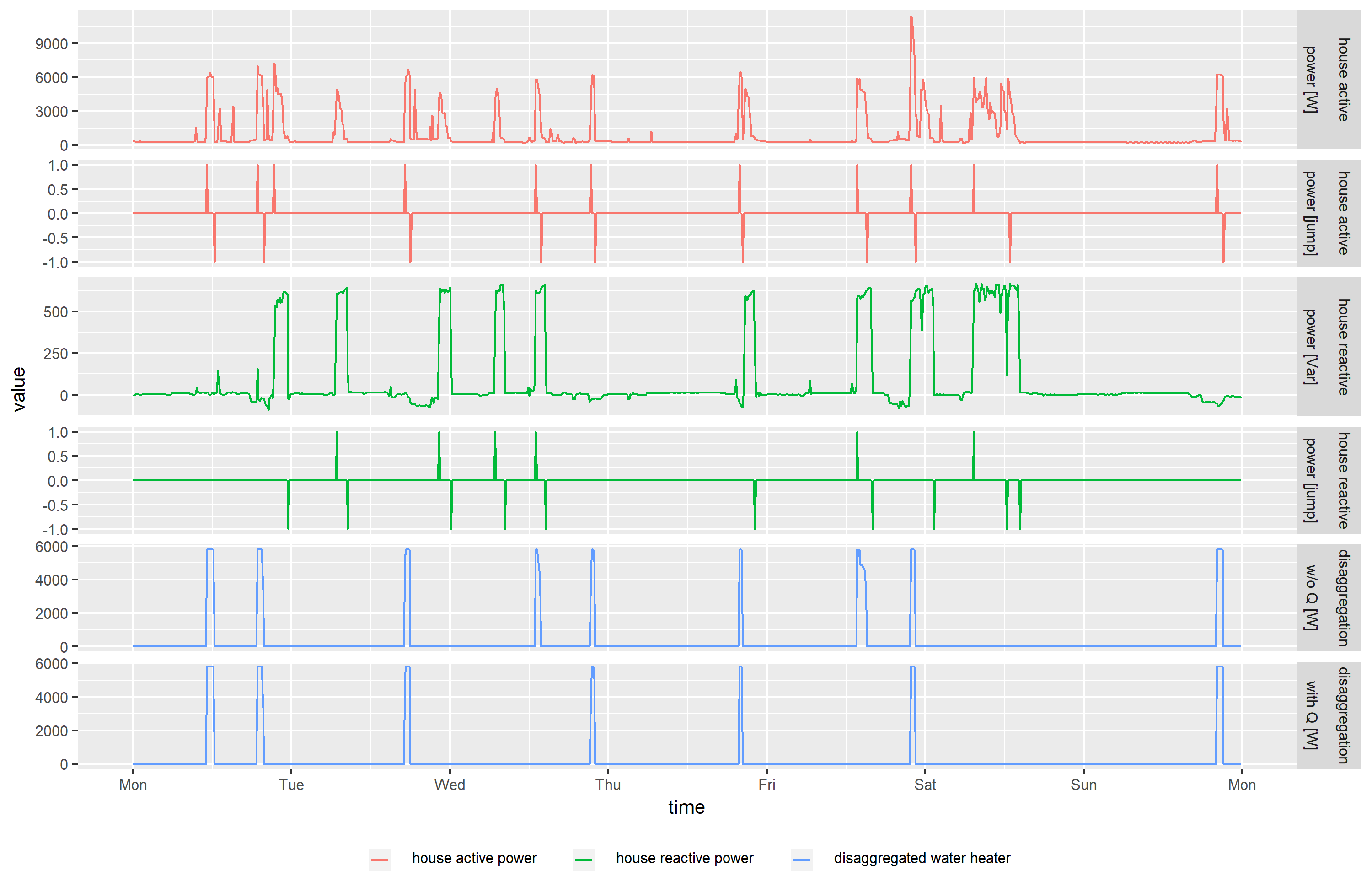}
\caption{Disaggregation process of a WH load based on the detection of jumps in the total active and reactive power profiles.}
\label{fig:process}
\end{figure*}

The core idea of the disaggregation process is the detection of large jumps in the total active and reactive power profiles. This concept is illustrated in Fig.~\ref{fig:process} for an example household (i.e.,~house 1 in Table~\ref{table:stats}) with 10-minute resolution data over 1 week. The first subplot represents the total active power profile of the household, from which upward and downward jumps in the same range as the estimated rated power are identified, corresponding to values of 1 and -1 in the second subplot. Note that the time resolution of the data is much lower than the transition between the ON and OFF periods of a WH which takes place almost instantaneously. Due to the time aggregation effect, the power level measured for the WH at the exact time step when the transition occurs is therefore a weighted average between zero and the rated power. Hence, a jump considers the difference in power over two consecutive time steps. If available, the same procedure is applied to the reactive power profile, as shown in the third and fourth subplots of Fig.~\ref{fig:process}. In this paper, the threshold for the identification of large jumps in active power is set $10\%$ below the rated power estimated from the total active power histogram as presented in Sec.~\ref{sec:detection}. Analogously, the threshold for large reactive power jumps is set $10\%$ below the estimated rated reactive power of the most recurrent inductive (or capacitive) load. This rated reactive power is estimated based on the histogram (bin width of 20 Var) of the total reactive power in absolute value. Note that the exact value of the power thresholds only marginally impacts the disaggregation as long as they are slightly lower than the estimated rated powers which are subject to small variations in reality.

As a further step, the identified jumps in the total active power profile are cleaned up to ensure an alternation of upward and downward jumps. Hence, ON periods of the water heater are defined between upward and downward jumps, and OFF periods are defined between downward and upward jumps. Moreover, only ON periods with a duration lower than 2 hours are retained, which is a realistic upper limit for typical water heaters. The resulting WH active load is zero during OFF periods and consists of the minimum between the estimated rated power and the total active power during ON periods:
\begin{equation}
    p^{\mathrm{WH}}_{t} =
    \begin{cases}
        0, \,  \mathrm{if} \; t \in \mathrm{OFF}\;\mathrm{period}, \\
        \min\left(\hat{p}^{\mathrm{WH}}_{r},p^{\mathrm{tot}}_{t}\right),\,  \mathrm{if} \; t \in \mathrm{ON}\;\mathrm{period},
    \end{cases}
\end{equation}
where $p^{\mathrm{WH}}_{t}$ is the estimated active power of the water heater at time $t$, $\hat{p}^{\mathrm{WH}}_{r}$ is the estimated rated power of the water heater, and $p^{\mathrm{tot}}_{t}$ is the active power of the total building at time $t$. As illustrated in the fifth subplot of Fig.~\ref{fig:process}, only part of the visible spikes in total active power are associated to the WH activities. The remaining spikes either have a too low maximum power, last longer than 2 hours, or require more than two time steps to reach their maximum. According to the proposed disaggregation approach, they likely correspond to other electric devices.

Finally, reactive power measurements can be leveraged to enhance the precision of the disaggregation process. Indeed, electric WHs are purely resistive devices (i.e.,~consuming only active power), in contrast to most other electric devices which are usually equipped with induction motors. Hence, the simultaneous detection of a jump in both active and reactive power profiles certainly reflects the activity of an inductive (or capacitive) electric device and cannot be attributed to a WH. In this way, the initial guess of the WH power profile can be further enhanced, where wrongly estimated ON periods are filtered out by the processing of the reactive power measurements. The comparison of the two last subplots of Fig.~\ref{fig:process} indicates that two ON periods are filtered out, corresponding to the activity of another electric device with reactive power consumption.

\section{Case Study}\label{sec:case_study}

The performance of the proposed detection and disaggregation approaches is evaluated on a real dataset with sub-metered WHs. The success of the disaggregation approach is first studied qualitatively before a quantitative analysis based on point-wise and classification metrics is carried out.

\subsection{Dataset}\label{sec:dataset}

To the best of our knowledge, no dataset with active and reactive power measurements of both the total building load and the WH load on a sufficiently large time period is publicly available. Hence, we have conducted our own power measurements at 1-minute resolution over two to four weeks in three different households of the City of San José, Costa Rica. Table~\ref{table:stats} details some specifications of the WHs in each household. Notably, they consume between one sixth and almost one third of the total household energy and the duration of their ON periods is extremely variable, which impacts the success of the proposed approach. The sub-metering of the WH load is solely used for validation purposes. Furthermore, 5-, 10-, and 15-minute resolution data, also typical of recent SM data, are created by averaging the 1-minute power values over the desired time granularity. Note that lower time resolutions (e.g., 30 and 60 minutes) are also common among SMs, however the residential power profiles get particularly smooth, which hinders the detection and the proper disaggregation of the WH load.

\begin{table}[!t]
 \caption{Specifications of water heaters used in the evaluation}
 \label{table:stats}
 \centering
 \begin{tabular}{|c|c|c|c|}
 \hline
    house id &
    \begin{tabular}{@{}c@{}}rated\\power [W]\end{tabular} & 
    \begin{tabular}{@{}c@{}}median ON/OFF\\duration [min]\end{tabular} &
    \begin{tabular}{@{}c@{}}share of\\energy [\%]\end{tabular} \\
    \hline
    1 & 6000 & 10 / 182 & 27 \\
    2 & 6000 & 3 / 193 & 17 \\
    3 & 3000 & 28 / 506 & 29 \\
 \hline
 \end{tabular}
\end{table}

\subsection{Evaluation Metrics}\label{sec:metrics}

The performance of the WH disaggregation approach is evaluated based on two different types of metrics. Error metrics focus on the point-wise error between the true and the estimated power values, whereas classification metrics assess whether the ON and OFF periods of the WH are estimated at the correct time steps.

\subsubsection{Error metrics}\label{sec:error_metric}

While the Mean Error (ME) makes the distinction between an overestimation and an underestimation of the actual power values, the Mean Absolute Error (MAE) gives the estimation error in absolute terms. Finally, the Normalized Mean Absolute Error (NMAE) allows fair comparison between multiple households with WHs of various rated powers. These three metrics are defined as follows:
\begin{subequations}
\begin{align}
    \text{ME} &:= \frac{1}{T}\sum_{t=1}^{T} \left(\hat{p}_t-p_t\right), \\
    \text{MAE} &:= \frac{1}{T}\sum_{t=1}^{T} |\hat{p}_t-p_t|, \\
    \text{NMAE} &:= 100\%\frac{1}{T \cdot p_r}\sum_{t=1}^{T} |\hat{p}_t-p_t|,
\end{align}
\end{subequations}
where $p_t$ and $\hat{p}_t$ are the true and estimated active power values at time $t$, respectively, $p_r$ is the rated power of the WH under consideration, and $T$ is the number of time steps. For error metrics, values close to zero indicate a good performance.

\subsubsection{Classification metrics}\label{sec:classification_metric}

In addition to the point-wise error, we also assess whether the time steps with and without WH activity are properly estimated, regardless of the power magnitude. Considering an event as a time step with WH consumption, the precision considers the portion of correct event estimations among all estimated events, the recall considers the portion of correct event estimations among all actual events, and the F1-score is the harmonic mean between the precision and the recall:
\begin{subequations}
\begin{align}
    \text{precision} &:= \frac{\text{true positives}}{\text{true positives} + \text{false positives}}, \\
    \text{recall} &:= \frac{\text{true positives}}{\text{true positives} + \text{false negatives}}, \\
    \text{F1-score} &:= 2\cdot\frac{\text{precision}\cdot\text{recall}}{\text{precision} + \text{recall}},
\end{align}
\end{subequations}
where true positives are correctly estimated events, false positives are incorrectly estimated events, and false negatives are true events which have not been estimated. For classification metrics, the higher the value, the better the performance.

\subsection{Performance Evaluation}\label{sec:evaluation}

\begin{figure*}[!t]
\includegraphics[width=1\linewidth]{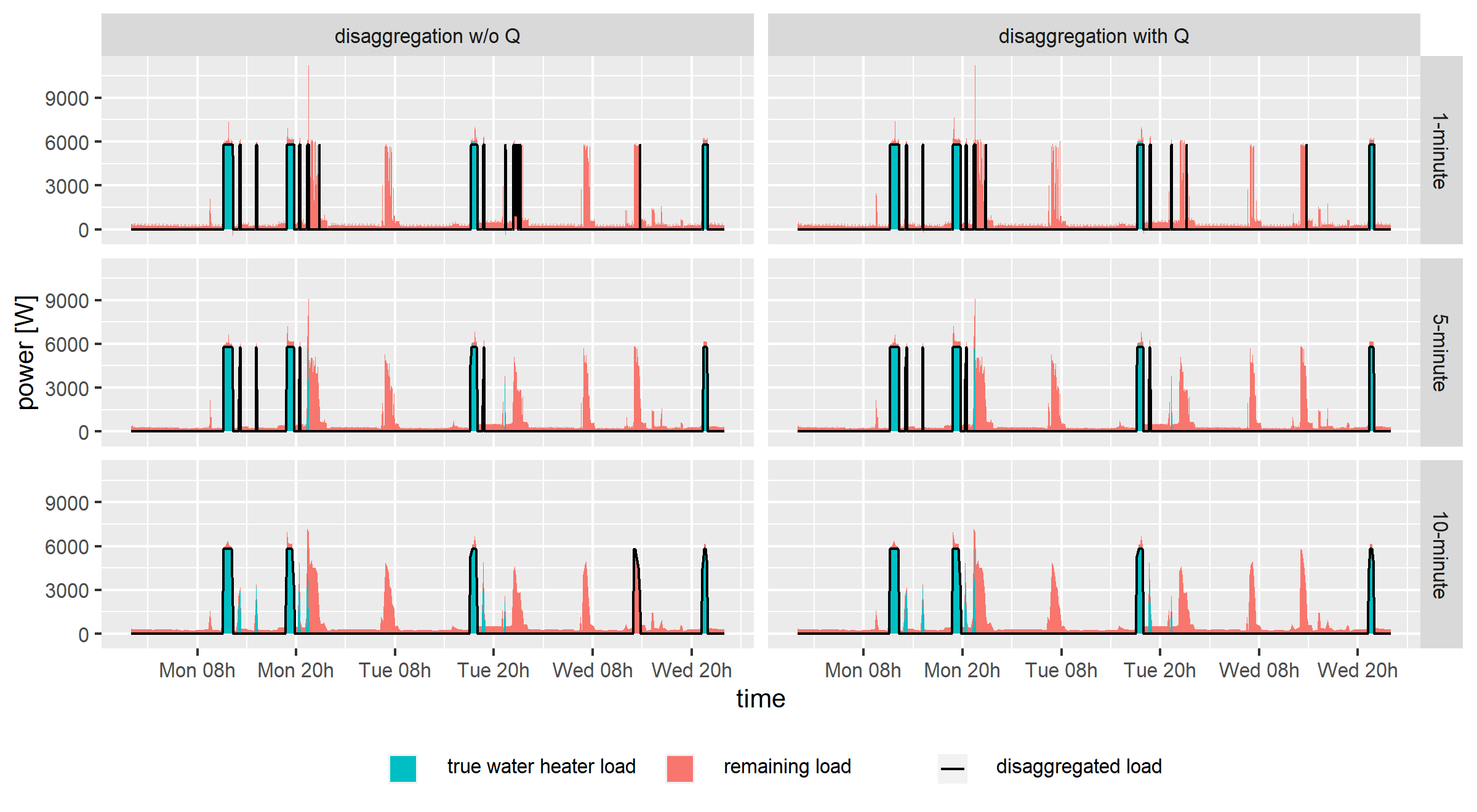}
\caption{Resulting disaggregation of water heater active power in an example household at different time resolutions.}
\label{fig:result}
\end{figure*}

\begin{figure}[!t]
\includegraphics[width=1\linewidth]{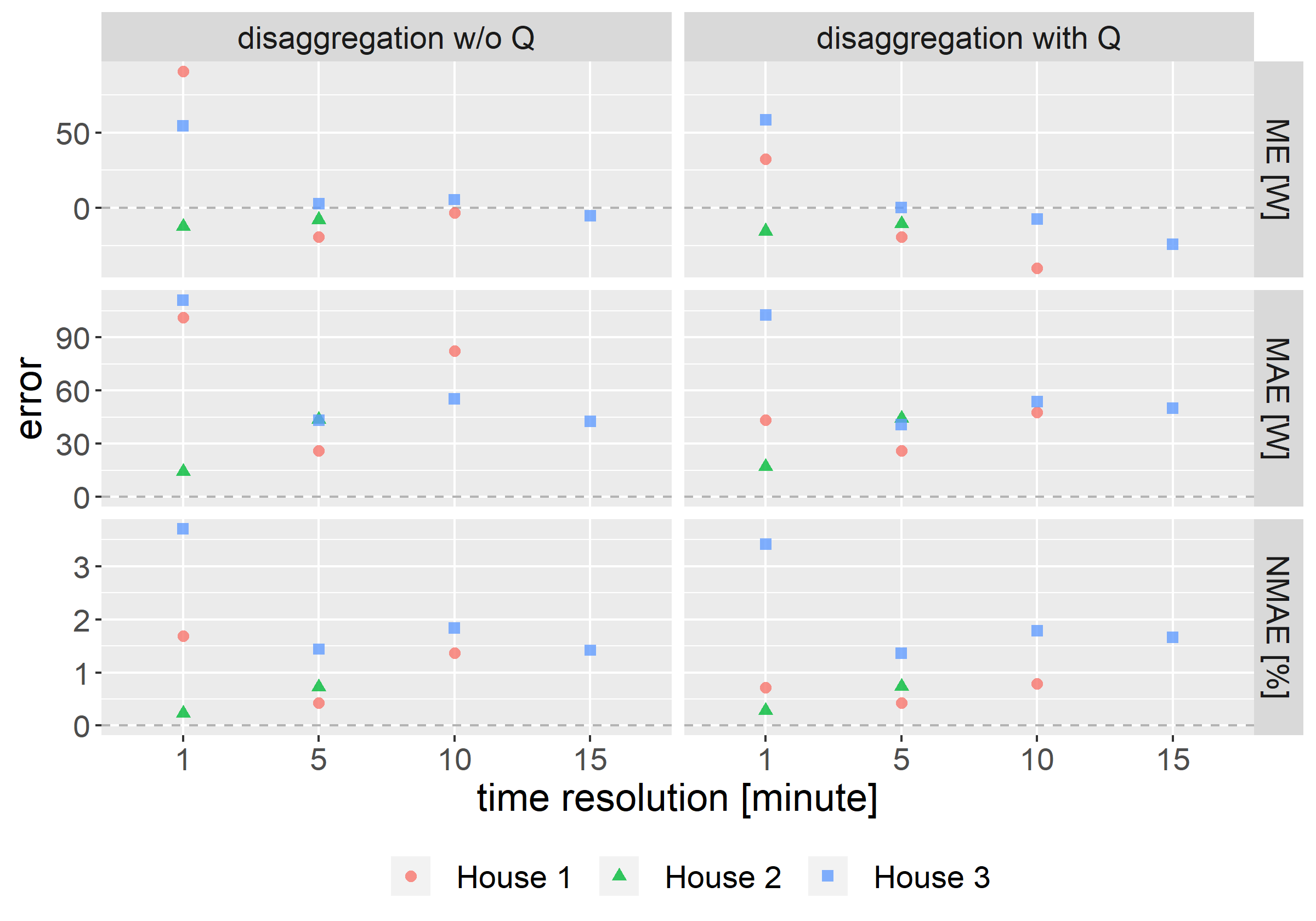}
\caption{Point-wise errors of the disaggregation of water heater loads}
\label{fig:error_metrics}
\end{figure}

\begin{figure}[!t]
\includegraphics[width=1\linewidth]{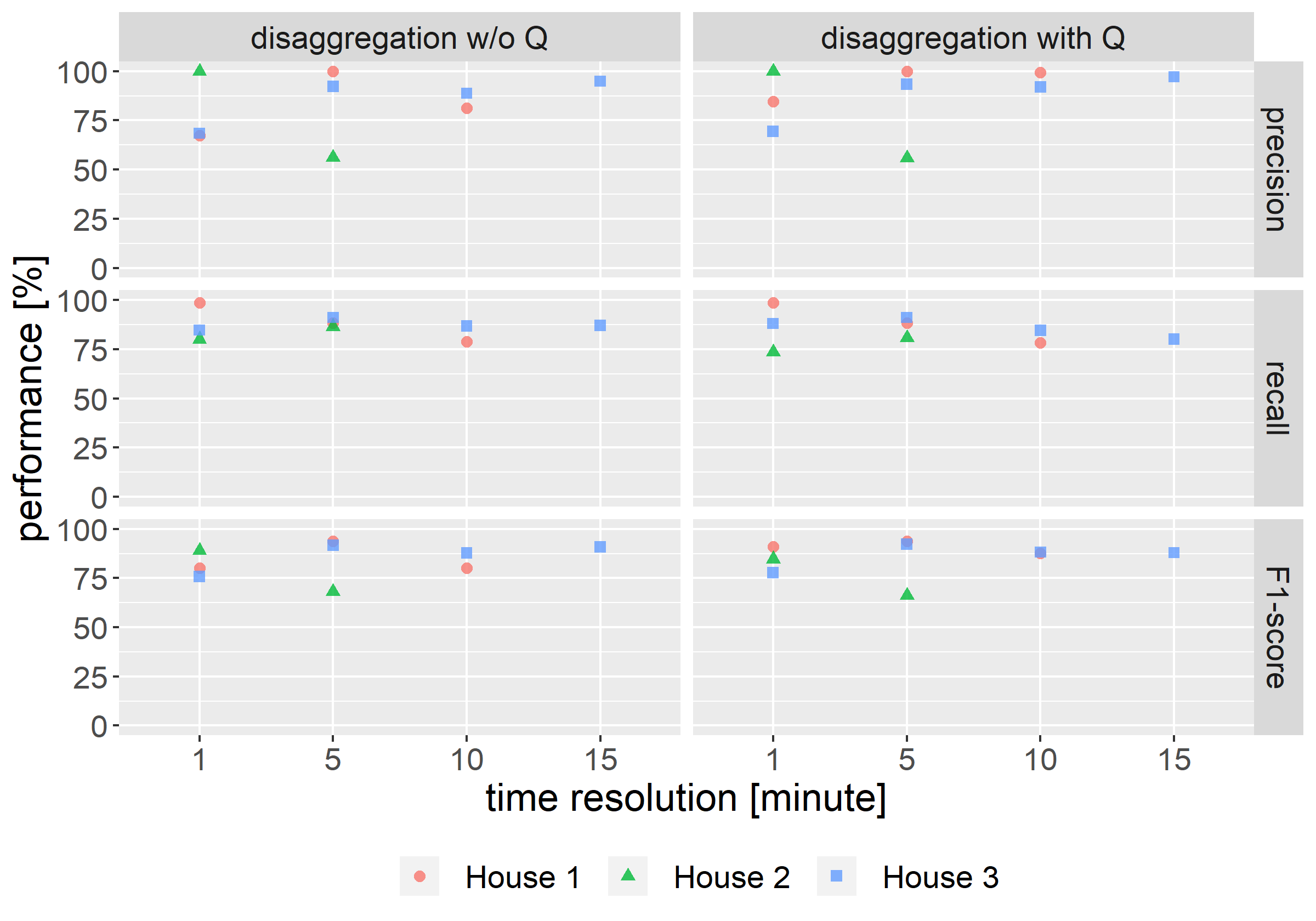}
\caption{Classification metrics of the disaggregation of water heater loads}
\label{fig:classification_metrics}
\end{figure}

Figure~\ref{fig:result} allows an intuitive interpretation of the disaggregation outcome by showing the true and estimated WH load profile over the three first days of Fig.~\ref{fig:process}. It is obvious that, in addition to the water heater with a rated power of 6 kW, another electric device consumes active power with a similar magnitude, which challenges the WH disaggregation approach. Furthermore, the time resolution impacts the shape of the total load profile and, consequently, alters the disaggregation outcome. The narrow and successive consumption spikes induced by the other large load observed at a 1-minute resolution (e.g.,~on Monday evening, Tuesday early morning, or Tuesday evening) are clearly smoothed out at lower time resolutions. At a 1-minute resolution, all WH events are correctly detected, but the proposed approach misclassifies some events of the other large load which shows similar features as the WH (false positive). The misclassification is however mitigated by the processing of reactive power measurements (Q). At a 10-minute resolution, the proposed approach still detects WH events with a long duration, but misses narrower WH events due to the smoothing effect which lowers the maximum observed power consumption (e.g., on Monday). Note that the use of reactive power measurements still allows the proposed approach to dispose of a false positive on Wednesday early afternoon.

Figures~\ref{fig:error_metrics} and~\ref{fig:classification_metrics} quantitatively illustrate the disaggregation performance for the three WHs specified in Table~\ref{table:stats}. First, the WH is not detected in house 1 with 15-minute data and in house 2 with 10- and 15-minute data. This is directly linked to the median ON duration of the WHs. When the ON duration is shorter than the time step duration, the averaged power value measured per time step is not the actual rated power, but depends on the actual ON duration. This also explains why the WH load of house 3, with a median ON duration of 28 minutes, is only impacted at a 30-minute resolution, as shown in the total load histogram in Fig.~\ref{fig:histogram}. Although the WH of house 2 is still detected at a 5-minute resolution, the smoothing effect already impacts a considerable part of the ON periods that cannot be identified any more, which translates into a substantial drop in precision and an increase of the point-wise errors due to overestimation.

Furthermore, the presence of highly volatile loads with a similar (or higher) power amplitude as the WH in houses 2 and 3 explains the relatively lower precision and higher error at 1-minute resolution, which is partially mitigated by the processing of reactive power measurements and is smoothed out at lower time resolutions. More generally, leveraging reactive power measurements during the disaggregation process typically increases the resulting precision, but sometimes at the cost of a slightly lower recall. Indeed, in rare cases, the WH is switched ON or OFF at the exact same time step as another inductive (or capacitive) load. Hence, by processing reactive power, the algorithm wrongly considers the jump in active power as a false positive.

Overall, the performance is still particularly high for houses and time resolutions for which a WH has indeed been detected, with a NMAE generally lower than $2\%$ (at the exception of house 3 with 1-minute data), a recall always higher than $73\%$, and a precision often close to $100\%$, especially when reactive power measurements are used. Finally, the algorithm tends to be conservative, which is reflected by a generally higher precision than recall and a ME slightly lower than zero, except at a 1-minute resolution.

\section{Conclusion}\label{sec:conclusion}

To summarize, this paper presents a novel approach to detect and disaggregate storage water heater loads from standard SM data without the need for sub-metering. WHs represent a large flexibility potential for DR schemes which can be greatly enhanced by an accurate estimation of the load profiles of individual heaters. While traditional NILM approaches are based on time resolutions and sometimes sub-metering data which are unrealistic in common distribution grids, the proposed approach only requires power measurements with a typical resolution between 1 and 15 minutes gathered by SMs at the customer level. It can easily be implemented either locally or in large distribution grids.

The presence of a water heater is first detected based on the total active power histogram, which is successful as long as the time resolution is not lower than the median ON duration of the water heater. Subsequently, the proposed algorithm detects and leverages features specific to water heaters in the total active power profile to extract the WH load profiles. Although the time resolution of the SM data plays a certain role, the disaggregation algorithm performs generally very good, showing an average NMAE of $1.3\%$ and an average F1-score of $85\%$ if a WH has initially been detected. If available, reactive power measurements can be similarly processed to enhance the precision of the disaggregation by discarding false positives.

In future work, a more robust evaluation of the proposed disaggregation approach requires a larger set of example houses with sub-metered WHs. Finally, a precise data-based estimation of the flexibility offered by the disaggregated water heaters for DR purposes on a local and a large scale will be carried out.

\end{document}